\documentclass[aps,prd,preprint,nofootinbib]{revtex4}
 \pdfoutput=1

\usepackage{amsmath,amssymb,amsfonts,latexsym,cancel,slashed}
\usepackage[mathscr]{euscript}

\usepackage{dcolumn}
\usepackage{graphicx}
\usepackage{color}

\def \bs {\boldsymbol}

\begin{document}
\title{Towards semi-inclusive deep inelastic scattering at next-to-next-to-leading order}
\author{Daniele Anderle $^{a,b}$, Daniel de Florian $^c$\footnote{On leave of absence from Departamento de F\'isica, FCEyN, Universidad de Buenos Aires} and Yamila Rotstein Habarnau $^{c}$}
\affiliation{$^a$ Institute for Theoretical Physics, University of T\"ubingen, Auf der Morgenstelle 14, 72076 T\"ubingen, Germany \\
$^b$ Lancaster-Manchester-Sheffield Consortium for Fundamental Physics, School of Physics
  \& Astronomy, University of Manchester, Manchester M13 9PL, United
  Kingdom\\
$^c$ International Center for Advanced Studies (ICAS), UNSAM, Campus Miguelete, 25 de Mayo y Francia, (1650) Buenos Aires, Argentina \\}

\begin{abstract}
In this paper, we compute the first set of ${\cal O}(\alpha_s^2)$ corrections to semi-inclusive deep inelastic scattering structure functions.
We start by studying the impact of the contribution of the partonic subprocesses that open at this order for the longitudinal
structure function. We perform the full calculation analytically, and obtain the expression of the factorized cross section at this order. 
 Special care is given to the study of their flavour decomposition structure.
We analyze the phenomenological effect of the corrections finding that, even though expected to be small a priori, it turns out to be sizable 
with respect to the previous order know, calling for a full NNLO calculation.
\end{abstract}

\maketitle

\section{Introduction}

Over the last decades, our understanding of hadron structure has remarkably improved, thanks to impressive experimental and theoretical progress. That includes the extraction of precise parton distribution functions (PDFs) from global analysis \cite{globalPDF}, complemented with accurate perturbative calculations for several processes in quantum chromodynamics (QCD). 
Recent progress has been observed towards a better description of the hadronization process, related experimentally to observables with identified light hadrons in the final-state. Their description relies on the previous two ingredients plus the knowledge of the corresponding fragmentation functions (FFs), which are evolving following the path of PDFs. It is in fact only recently that a first next-to-next-to-leading order (NNLO) analysis of FFs based on electron-positron annihilation data was presented in \cite{Anderle:2015lqa}. A global analysis including also proton-proton collision's data and semi-inclusive deep inelastic scattering (SIDIS) data at this precision level is still yet to come. Therefore, the computation of NNLO corrections to the SIDIS process is an absolute requirement in order to extend existing NLO global analyses \cite{deFlorian:2014xna,FFs}.
Analyses solely based on electron-positron annihilation into hadrons give no information on how the individual quark flavour fragments into hadrons, and leave a 
considerable uncertainty on the gluon density.
The measurement of final state hadrons in SIDIS provides an excellent complementary tool for the extraction of fragmentation functions. Furthermore, SIDIS plays a very important role in understanding the spin structure of the nucleon, that can be described by the (non-perturbative) polarized parton distribution functions (pPDFs). The most complete global  fit of pPDFs includes all available data taken in spin-dependent DIS, semi-inclusive DIS with identified pions and kaons, and proton-proton collisions. These fits allow the extraction of pPDFs consistently at NLO \cite{deFlorian:2014yva}. In particular in this context, SIDIS with identified hadrons in the final state is of essential need in order to achieve a full flavour decomposition for the polarized parton distributions.

For all these reasons, counting on precision theoretical description for SIDIS is mandatory.
In the fully-inclusive case the structure functions are well known at next-to-next-to-leading order (NNLO) in perturbative QCD, both for the unpolarized \cite{vanNeerven:1991nn,Zijlstra:1992qd,Moch:1999eb} and for the polarized ones \cite{Zijlstra:1993sh}. 
For the unpolarized case, even the hard corrections at order $\mathcal{O}(\alpha_s^3)$ are available \cite{Vermaseren:2005qc}. 
However, for semi-inclusive DIS, the QCD corrections are only known up to NLO both in the unpolarized \cite{Altarelli:1979kv,deFlorian:1997zj} and the polarized cases \cite{deFlorian:1997zj}.

Nowadays, NNLO is the state of the art for many observables of interest. It is then natural to try to reach the same accuracy for the unpolarized SIDIS process. In an effort to analytically calculate corrections at this level of precision, one may start by analyzing the simpler case of the longitudinal component of the process, in order then to use the acquired experience to extend the calculation to the more difficult transverse one. Both components are essential to evaluate the ratio between the longitudinal and transverse photoabsortion cross sections $R \equiv {\sigma_L}/{\sigma_T} $, which plays an important role in the extraction of pPDFs from the observed asymmetries (see for instance \cite{Alekseev:2010ub}). In such analyses, the semi-inclusive ratio $R$ is customarily assumed to be equal to the inclusive one, which may not be always a good approximation.

In this paper we perform the first steps towards the computation of the longitudinal SIDIS structure function at NNLO accuracy. In particular, we focus on the contribution of those channels that open for the first time at this order. In section \ref{sec:SIDIS} we introduce the SIDIS structure functions and the cross section ratio. Their flavour decomposition structure is discussed in section \ref{sec:NNLO}. In section \ref{sec:FL_NNLO} we explain the main features of the computation of the new contributions to the longitudinal structure function at NNLO. In section \ref{sec:Results} we present some phenomenological results and finally the conclusions are presented in section \ref{sec:Conclusions}.

\section{Semi-inclusive deep inelastic scattering}
\label{sec:SIDIS}
%

The cross section for the scattering of leptons on nucleons with the observation of a hadron $H$ in the final state can be written, in lowest-order perturbation theory of electroweak interactions, as 
\begin{equation}
\frac{d\sigma^H}{dx\,dy\,dz}=\frac{2\pi\, y \, \alpha^2}{Q^4} \sum_{j}  L^{\mu\nu}W^{H}_{\mu\nu},
\end{equation}
where  the leptonic tensor $L^{\mu\nu}$ is associated with the coupling of the exchanged photon  to the leptons  (we do not include processes mediated by $Z$ and $W$ bosons) while the hadronic tensor $W_{\mu\nu}^H$ describes the interaction of the photon with the target nucleon and the hadronization of partons into $H$. 
Here, $x$ and $y$ denote the usual DIS variables:
\begin{equation}
-q^2=Q^2=S x y, \quad  \quad  x=Q^2/(2P\cdot q), \nonumber
\end{equation}
where $q$ is the photon four-momentum, $P$ the nucleon momentum and $S$ the center-of-mass energy squared of the lepton-nucleon system. Besides, $z=P_H\cdot P/P\cdot q$ is the scaling variable representing the momentum fraction taken by the hadron $H$. Since we concentrate in the current fragmentation region, cuts over $z$ should apply (typically, $z>0.1$)\footnote{Due to the definition of $z$, the target fragmentation process \cite{Daleo:2003xg,Daleo:2003jf} is strictly $z=0$.}.

The unpolarized SIDIS structure functions ($F_i^H$) are defined in terms of the hadronic tensor. Besides terms that cancel after integrating over the azhimutal angle of the outgoing hadron, one gets the usual DIS tensor: \cite{Levelt:1993ac}
\begin{align}
W^{H}_{\mu\nu}=&\left( -g_{\mu\nu}+\frac{q_{\mu}q_{\nu}}{q^2}\right) \; F_1^{H}(x,z,Q^2) 
		 + \frac{\hat{P}_\mu \, \hat{P}_\nu}{P\cdot q} \; F_2^{H}(x,z,Q^2) ,
\label{eq:Wmunu}
\end{align}
where ${\hat{P}_{\mu}= P_{\mu}-\frac{P\cdot q}{q^2}\;q_{\mu}}$. We have not taken into account those terms proportional to the polarized structure functions.

The spin-averaged SIDIS cross section for $Q^2\gg M^2$ ($M$ being the mass of the target nucleon), is then given by
\begin{equation}
\frac{d\sigma^{H}}{dx \,dy \,dz} = \frac{2\pi\alpha^2}{x\,y\,Q^2}  \Bigg[
\big[ 1+(1-y)^2 \big]\, 2x\, F_1^{H}  +(1-y) \, 2 \, F_L^{H} \Bigg] \,\, .
\label{eq:sigma_nopol}
\end{equation}
The longitudinal structure function is defined as $F_L^H=F_2^H-2\, x\, F_1^H$ and 
vanishes at lowest order \cite{Callan:1969uq}.

Defining the ratio
\begin{equation}
R^H=\frac{\sigma_L^H}{\sigma_T^H}=\frac{F_L^H}{2\, x\, F_1^H} \,,
\label{eq:R}
\end{equation}
where $\sigma_L^H$ and $\sigma_T^H$ are the semi-inclusive cross section for longitudinal and transversely polarized virtual photons respectively, Eq. \eqref{eq:sigma_nopol} can be rewritten as
\begin{equation}
\frac{d\sigma^{H}}{dx \,dy \,dz} = \frac{2\pi\alpha^2}{x\,y\,Q^2}  \, F_2^H \, \Bigg[
\big[ 2(1-y) + \frac{y^2}{1+R^H} \Bigg] \,\, .
\label{eq:sigma_nopol_R}
\end{equation}

\section{The structure functions at next-to-next-to leading order}
\label{sec:NNLO}
%
Assuming factorization, the SIDIS structure functions can be obtained as the convolution of parton distribution functions (PDFs) and fragmentation functions (FFs), describing the low-energy non perturbative behaviour, with 
short-distance coefficients that can be evaluated in perturbation theory. In general we can write 
\begin{align}\label{eq:strucfunc}
F_{k}(x,z,Q^2,\mu^2_F,\mu^2_I,\mu_r^2)=&\bigg[\sum_{q_a,q_b} q_a\otimes C_{k}^{q_a,q_b}\otimes D^h_{q_b}+\sum_{q_a} q_a\otimes C_{k}^{q_a,g}\otimes D^h_g \nonumber \\
&+\sum_{q_b} g \otimes C_{k}^{g,q_b}\otimes D^h_{q_b}+g\otimes C_{k}^{g,g}\otimes D^h_{g}\bigg](x,z,Q^2,\mu^2_F,\mu^2_I,\mu_r^2)\;,
\end{align}
where $k\in\{1,L\}$, the sums are understood to run over all possible quark and anti-quark flavours and $\otimes$ denotes the usual convolution,
\begin{align}\label{eq:convolution}
(q \otimes C \otimes D^h)(x,z,Q^2,\mu^2_I,\mu_F^2,\mu_r^2) =& \int_x^1\frac{dy}{y}  \int_{z}^1 \frac{d\omega}{\omega}  \, q(y,\mu_I^2)C\left(\frac{x}{y},\frac{z}{\omega},\mu_r^2,\frac{Q^2}{\mu_I^2},\frac{Q^2}{\mu_F^2},\frac{Q^2}{\mu_r^2}\right) \nonumber \\
& \times D^h(\omega,\mu_F^2) \, \,.
\end{align}
The coefficient functions $C_{k}^{ij}$  (with $i$ and $j$ denoting the initial and hadronizing partons respectively) can be perturbatively calculated as a series in the strong coupling constant $\alpha_s$,
\begin{equation}\label{eq:CExpansion}
C_{k}^{ij}\left(x,z,\mu_r^2,\frac{Q^2}{\mu_I^2},\frac{Q^2}{\mu_F^2},\frac{Q^2}{\mu_r^2}\right)=\sum_{n} \left(\frac{\alpha_s(\mu_r^2)}{4\pi}\right)^n C_{k}^{ij\;(n)}\left(x,z,\frac{Q^2}{\mu_I^2},\frac{Q^2}{\mu_F^2},\frac{Q^2}{\mu_r^2}\right)\;.
\end{equation}
The renormalization scale $\mu_r$ represents the ``hard-scale'' at which the perturbative expansion is performed whereas the factorization scales $\mu_I$ and $\mu_F$ separate conceptually the perturbative regime from the non-perturbative one in the initial and final state part of the process respectively. The PDFs $q$ and $g$, describing the momentum fraction distribution of the parton inside the struck hadron, and the FFs $D^h_q$ and $D^h_g$, describing the fragmentation of the parton into an hadron $h$, are process independent distributions that can be extracted from data through global QCD analysis of reference processes. Although they cannot be obtained from first principles in perturbative QCD, it is possible to predict their dependence on the factorization scale $\mu_{I,F}$ once they are given at some reference scale $\mu_0$ by solving the Dokshitzer-Gribov-Lipatov-Altarelli-Parisi (DGLAP) evolution equations \cite{DGLAP}. Their respective space-like and time-like versions read
\begin{align}\label{eq:DGLAPspace}
\frac{\partial}{\partial\log\mu_I^2} f_i(x,\mu_I^2)&=\sum_j \left(P_{ij}\left(\mu_r^2,\frac{\mu_I^2}{\mu_r^2}\right)\otimes f_j(\mu_I^2)\right)(x)\\\label{eq:DGLAPtime}
\frac{\partial}{\partial\log\mu_F^2} D^h_{f_i}(z,\mu_F^2)&=\sum_j \left(P^T_{ji}\left(\mu_r^2,\frac{\mu_F^2}{\mu_r^2}\right)\otimes D^h_{f_j}(\mu_F^2)\right)(z)\;.
\end{align}
Here the sum runs over all possible $f_j = q_j,\;\bar q_j,\; g$. The space-like and time-like splitting functions, $P_{ij}$ and $P^T_{ji}$ respectively, are perturbative calculable functions. In the space-like case, for example, the expansion in $\alpha_s(\mu_r)$ can be written as
\begin{align}\label{eq:Pexpansion}
P_{ij}\left(x,\mu_r^2,\frac{\mu_r^2}{\mu_I^2}\right)=&\sum_{n} a_s^{n+1}\left(\mu_r^2,\frac{\mu_r^2}{\mu_I^2}\right) P_{ij}^{(n)}(x)\nonumber\\
=&a_s(\mu_r^2)P_{ij}^{(0)}(x)+a_s^2(\mu_r^2)\left(P_{ij}^{(1)}(x)+\beta_0 P_{ij}^{(0)}(x)\log\left(\frac{\mu_r^2}{\mu_I^2}\right)\right)\nonumber\\
&+a_s^3(\mu_r^2)\left(P_{ij}^{(2)}(x)+2\beta_0 P_{ij}^{(1)}(x)\log\left(\frac{\mu_r^2}{\mu_I^2}\right) \right. \nonumber \\ &\left. +\left\{  \beta_1 \log\left(\frac{\mu_r^2}{\mu_I^2}\right)+  \beta_0^2 \log^2\left(\frac{\mu_r^2}{\mu_I^2}\right)\right\}P_{ij}^{(0)}(x)\right)+\dots\nonumber\\
=&\sum_n a_s^{n+1}(\mu_r^2) \left (P_{ij}^{(n)}(x)+ \sum_{m=1}^{n}\log^m\left(\frac{\mu_r^2}{\mu_I^2}\right)\sum_{k=0}^{n-m} A^{n+1}_{k,\,m}P_{ij}^{k}(x)\right)\;,
\end{align}
where $a_s=\alpha_s/4\pi$ and $\beta_i$ are the usual expansion coefficients of the QCD beta function $\beta(a_s) = - a_s^2 \sum_{i=0}^\infty \beta_i \,a_s^i$. The second equality was obtained by re-expanding $a_s({\mu_r^2},\mu_r^2/\mu_I^2)$ in terms of $a_s({\mu_r^2})$ (see Eq.~\eqref{eq:expansion}). The coefficients $A^{n+1}_{k,\,m}$ collect the beta terms coming from this expansion and they will be of use for further discussion in Appendix \ref{sec:scales}. For the sake of notation and simplicity, we can proceed by setting all scales equal, $\mu_I^2=\mu_F^2=\mu_r^2=Q^2$ without loss of information. A sketch on how it is possible to recover all scale dependences is given in Appendix \ref{sec:scales} for a specific case.

Eqs.~\eqref{eq:DGLAPspace}  for the PDFs and~\eqref{eq:DGLAPtime} for the FFs are each $2N_f+1$ integro-differential coupled equations, with $N_f$ being the number of active massless flavours. It is customary to rewrite the quark sector into flavour singlet combinations
\begin{equation}\label{eq:singlet}
q_S\equiv\frac{1}{N_f}\sum_{i}^{N_f}(q_i+\bar q_i),~~~~~~~~D^h_S\equiv\frac{1}{N_f}\sum_{i}^{N_f}(D^h_{q_i}+ D^h_{\bar q_i}),
\end{equation}
which evolve together with $g$ and $D^h_g$ respectively according to 
\begin{equation}\label{eq:DGLAPsinglet}
\frac{\partial}{\partial \log{Q^2}}\Bigg (\begin{matrix} q_S\\[2mm] g \end{matrix} \Bigg )=
\Bigg (\begin{matrix}P_{qq} & P_{qg}\\[2mm] P_{gq} & P_{gg} \end{matrix} \Bigg )\otimes\Bigg (\begin{matrix} q_S\\[2mm] g \end{matrix} \Bigg ),~~~~~~~~\frac{\partial}{\partial \log{Q^2}}\Bigg (\begin{matrix} D^h_{S}\\[2mm] D^h_g \end{matrix} \Bigg )=
\Bigg (\begin{matrix}P^T_{qq} & P^T_{gq}\\[2mm] P^T_{qg} & P^T_{gg} \end{matrix} \Bigg )\otimes\Bigg (\begin{matrix} D^h_{S}\\[2mm] D^h_g \end{matrix} \Bigg )\,,
\end{equation}
and three non-singlet combinations for PDFs and for FFs
\begin{align}\label{eq:nonsinglet}
q_{ns,\,ik}^{\pm}&=q_i\pm\bar q_i-(q_k\pm \bar q_k)  &D_{ns,\,ik}^{h,\,\pm}&=D^{h}_{q_i}\pm D^{h}_{q_i}-(D^h_{q_k}\pm D^h_{\bar q_k})\\
q_{i}^{v}&=q_i-\bar q_i &D_{q_{i}}^{h,\,v}&=D^h_{q_i}-D^h_{\bar q_i}
\end{align}
which evolve independently with $P^{+}_{ns}$, $P^{-}_{ns}$, $P^{v}_{ns}$, $P^{T,\,+}_{ns}$, $P^{T,\,-}_{ns}$, $P^{T,\,v}_{ns}$ and decouple the remaining $2N_f-1$ equations. All splitting functions are known up to NNLO \cite{Moch:2004pa,Vogt:2004mw,Mitov:2006ic,Moch:2007tx,Almasy:2011eq}.

\begin{figure}[t]
\vspace*{-0.4cm}
\begin{center}
\includegraphics[width=0.9\textwidth]{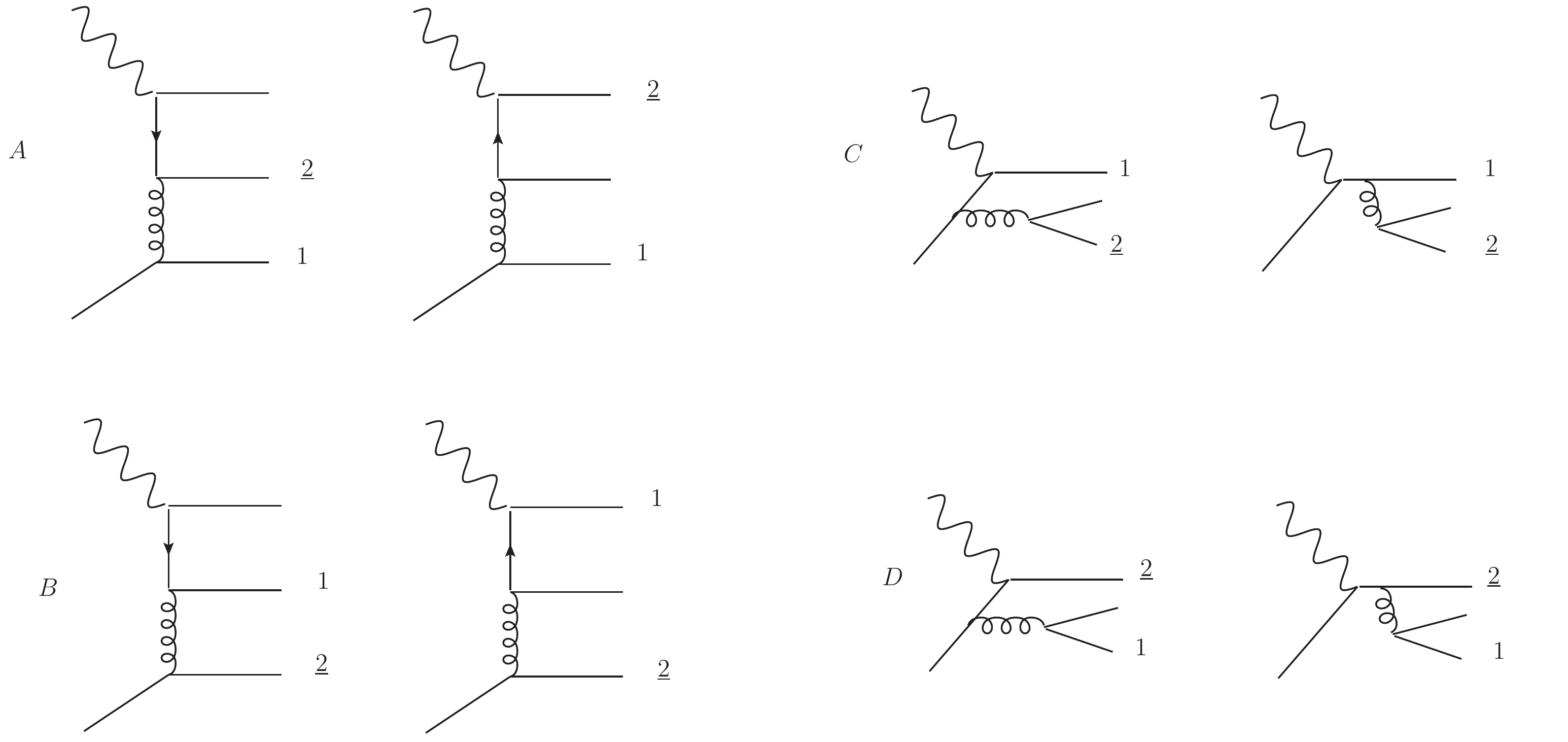}
\end{center}
\caption{Diagram contributions to the sub-process $\gamma*+q\rightarrow q(1)+``q"(2)+\bar q $ (and $\gamma*+\bar q\rightarrow \bar q(1)+ ``\bar q\,"(2)+ q $ if the arrows are inverted in group $A$ and $B$). Particle ``2'' is assumed to be the one fragmenting in the semi-inclusive case. 
\label{fig:qqNNLO} }
\end{figure}

As it is done in the literature for the totally inclusive case~\cite{Zijlstra:1992qd,Furmanski:1981cw}, the structure functions in Eq.~\eqref{eq:strucfunc} can be explicitly written as functions of non-singlet and singlet PDFs and FFs combinations. This is especially relevant at NNLO since different diagrammatic contributions to the flavour combinations are made manifest. In the DIS inclusive case, for example, it is common to write the structure functions separating the ``non-singlet'' (NS) from the ``singlet'' (S) contributions $C_{k}^{\text{NS}}$ and $C_{k}^{\text{S}}$ which at $\mathcal{O}(a^2_s)$ start to differ from each other \cite{Zijlstra:1992qd}:
\begin{align}\label{eq:DISstrucfunc}
F^{\text{DIS}}_{k}(x,Q^2)&=\sum_{j}  \left (C_{k}^{\text{DIS},\,q_j}\otimes q_{j}+C_{k}^{\text{DIS},\,\bar q_j}\otimes \bar q_{j}\right)+C_{k}^{\text{DIS},\,g}\otimes g\\\label{eq:DISstrucfunc2}
&=\sum_{j} e^2_{q_j} C_{k}^{\text{NS}}\otimes q^{\text{NS}}_j(x,Q^2)+\bigg(\sum_{j} e^2_{q_j}  \bigg)\left[   C_{k}^{\text{S}} \otimes q_S +  C_{k}^{\text{DIS},\,g} \otimes g\right](x,Q^2)\;,
\end{align}
where $k\in\{1,L\}$, $e_{q_j}$ are the electromagnetic charges of quarks and all sums run over the active flavours. The flavour combination $q^{\text{NS}}_j$ is defined as
\begin{equation}\label{eq:qNS}
q^{\text{NS}}_j\equiv \frac{1}{N_f} \sum_{k=1}^{N_f} q_{ns,\,jk}^{+} =(q_j+\bar q_j) - \frac{1}{N_f}\sum_{k=1}^{N_f}(q_k+\bar q_k)
\end{equation}
and therefore evolves with $P^+_{ns}$ whereas $q_S$ was defined in~\eqref{eq:singlet} and evolves according to~\eqref{eq:DGLAPspace}. The equality between \eqref{eq:DISstrucfunc} and \eqref{eq:DISstrucfunc2} is a direct consequence of the charge conjugation symmetry $C_{k}^{\text{DIS},\,q_i}=C_{k}^{\text{DIS},\,\bar q_i}$ when the considered incoming vector is a photon. In fact we can distinguish between NS diagrammatic contributions and ``pure-singlet'' (PS) ones and write
\begin{equation}\label{eq:coefDISps}
C_{k}^{\text{DIS},\,q_i}=C_{k}^{\text{DIS},\,\bar q_i}=e_{q_i}^2 C_{k}^{\text{NS}}+\frac{1}{N_f}\bigg(\sum_{j} e^2_{q_j}  \bigg) C_{k}^{\text{PS}}\;. 
\end{equation}
In this case one defines NS contributions to be the ones where either on the left side or on the right side of the cut diagrams the struck parton is directly connected to the incoming quark through a quark line (e.g. at NNLO $C^2$ or $BC$ in Fig.~\ref{fig:qqNNLO}). On the other hand, PS contributions generate from cut diagrams where on both sides of the cut the struck parton is separated by gluon lines from the incoming quark (e.g. at NNLO $A^2$ in Fig.~\ref{fig:qqNNLO}). 
Inserting Eq.~\eqref{eq:qNS} in~\eqref{eq:DISstrucfunc} one obtains ~\eqref{eq:DISstrucfunc2} with $C_{k}^{\text{S}}=C_{k}^{\text{NS}}+C_{k}^{\text{PS}}$. Charge conjugation breaking terms proportional to $e_i\,\sum_j e_j$ vanish at each order either due to their colour structure or due to Furry's theorem, which means that~\eqref{eq:coefDISps}, and therefore~\eqref{eq:DISstrucfunc2}, are all-order valid equalities.

\begin{figure}[t]
\vspace*{-0.4cm}
\begin{center}
\includegraphics[width=0.9\textwidth]{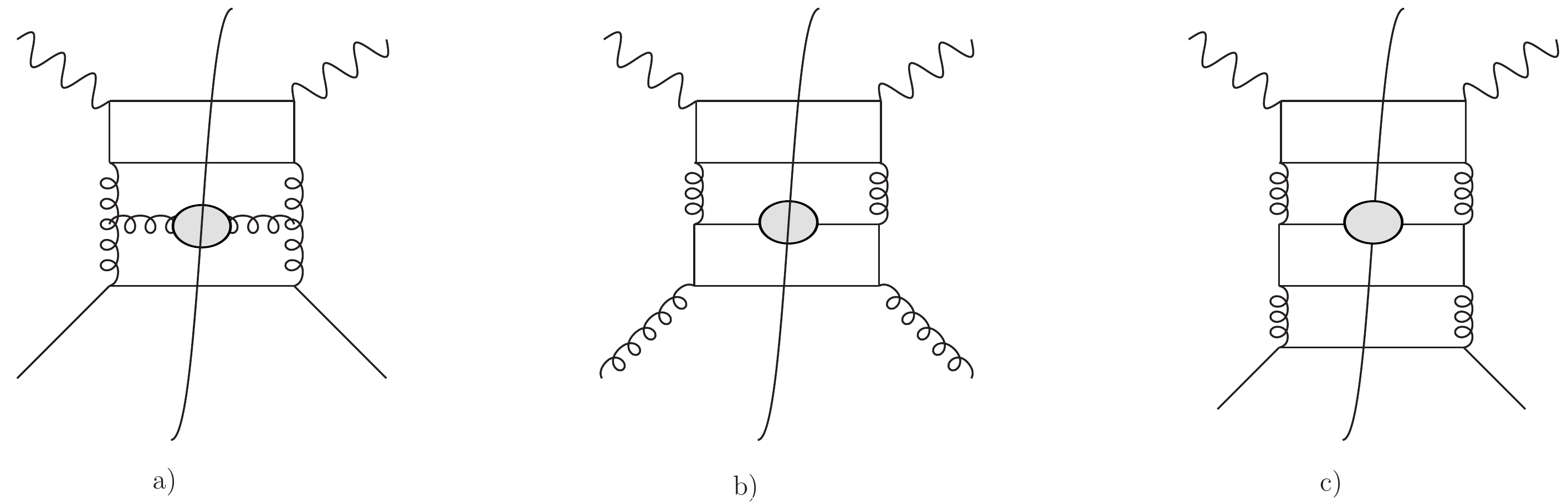}
\end{center}
\caption{Cut diagrams proportional to $\sum_i\left(e^2_{q_i}\right)/N_f$. The grey blob indicates the fragmenting outgoing particle. a), b) contribute to the third order $C^{q_i,g,\,(3)}$ and $C^{g,q_i,\,(3)}$ respectively whereas c) contributes both to the fourth order $C^{q_i,\bar q_i,\,(4)}$ and to $C^{q_i,q_j,\,(4)}$.
\label{fig:NmNLO} }
\end{figure}

In the semi-inclusive case, the identification of a final state hadron complicates the above described diagrammatic contribution's separation since $C_{k}^{q_i,\, q_i}\neq C_{k}^{q_i,\, \bar q_i}\neq C_{k}^{q_i,\, q_j}$. In particular non vanishing terms proportional to $e_i e_j$ start to appear at NNLO due to contributions where the convolution with different FFs for quark and anti-quark spoils Furry's theorem: for example the $q_1\otimes C_{k}^{q_1,\, q_2,\;(AC)}\otimes D^h_{q_2}$ and $q_1\otimes C_{k}^{q_1,\, \bar q_2,\;(AC)}\otimes D^h_{\bar q_2}$ terms generating from the interference term AC in Fig~\ref{fig:qqNNLO} do not vanish in the sum since in general $D^h_{q_2}\neq D^h_{\bar q_2}$ although $C_{k}^{q_1,\, q_2,\;(AC)}=-C_{k}^{q_1,\, \bar q_2,\;(AC)}$. By introducing the corresponding  time-like ``non-singlet'' combinations
\begin{equation}\label{eq:DqNS}
D_{q_j}^{h,\,\text{NS}}\equiv \frac{1}{N_f} \sum_{k=1}^{N_f} D_{ns,\,ik}^{h,\,+}=(D^h_{q_j}+D^h_{\bar q_j}) - \frac{1}{N_f}\sum_{k=1}^{N_f}(D^h_{q_k}+D^h_{\bar q_k})\;,
\end{equation}
we can express the semi-inclusive structure functions~\eqref{eq:strucfunc} as
\begin{eqnarray}\label{eq:SIDSstrucfunctFlav}
F&=&(q_S,  g )\otimes \Bigg(\begin{matrix}  \mathcal{C}^{S,D_S}& \mathcal{C}^{S,g} \\[2mm] \mathcal{C}^{g,D_S}& \mathcal{C}^{g,g}\end{matrix}\Bigg)\otimes \Bigg(\begin{matrix}  D^h_S \\[2mm] D^h_g\end{matrix}\Bigg)
\nonumber\\[2mm]
&&+\sum_{i}^{N_f}q_i^{\text{NS}}\otimes (\mathcal{C}^{\text{NS},D_S}_{q_i},\mathcal{C}^{q_i,g})\otimes \Bigg(\begin{matrix}  D^h_S \\[2mm] D^h_g\end{matrix}\Bigg)
+\sum_{j}^{N_f} (q_S,  g )\otimes\Bigg(\begin{matrix}  \mathcal{C}^{S,D_\text{NS}}_{q_j} \\[2mm] \mathcal{C}^{g,q_j}\end{matrix}\Bigg)\otimes D_{q_j}^{h,\,\text{NS}}
\nonumber\\[2mm]
&&+\sum_{i,j}^{N_f}q_i^{\text{NS}}\otimes \mathcal{C}^{\text{NS}}_{q_i,q_j}\otimes D_{q_j}^{h,\,\text{NS}}
+\sum_{i}^{N_f}q_i^{v}\otimes \mathcal{C}^{\text{v}}_{q_i,q_i}\otimes D_{q_i}^{h,\,v}\nonumber\\[2mm]
&&+\sum_{\substack{i,j\\i\neq j}}q_i^{v}\otimes \mathcal{C}^{\text{v}}_{q_i,q_j}\otimes D_{q_j}^{h,\,v}
\end{eqnarray}
where the index ``$k$'' and the dependencies on $x$, $z$ and $Q^2$ were dropped in order to simplify the notation. The above formula is valid to  all orders and the different coefficient functions $\mathcal{C}$ relate to the coefficient functions $C$ in \eqref{eq:CExpansion} according to the following equalities:
\begin{align}\label{eq:CoefRelation}
&\mathcal{C}^{S,D_S}=\frac{1}{2}\sum_{i}^{N_f}\sum_{j}^{N_f}\left(C^{q_i,q_j} + C^{q_i,\bar q_j}\right)  &       
& \mathcal{C}^{S,g}=\sum_{i}^{N_f} C^{q_i,g}
~~~~\mathcal{C}^{g,D_S}=\sum_{j}^{N_f} C^{g,q_j} &
& \mathcal{C}^{g,g}=C^{g,g} 
\nonumber\\[2mm]
&\mathcal{C}^{\text{NS},D_S}_{q_i}=\frac{1}{2}\sum_{j}^{N_f}\left(C^{q_i,q_j} + C^{q_i,\bar q_j}\right) &
& \mathcal{C}^{q_i,g} =C^{q_i,g} 
\nonumber\\[2mm]
& \mathcal{C}^{S,D_\text{NS}}_{q_j} =\frac{1}{2}\sum_{i}^{N_f}\left(C^{q_i,q_j} + C^{q_i,\bar q_j}\right)&
& \mathcal{C}^{g,q_j} =C^{g,q_j}&
\nonumber\\[2mm]
& \mathcal{C}^{\text{NS}}_{q_i,q_j} =\frac{1}{2}\left(C^{q_i,q_j} + C^{q_i,\bar q_j}\right)&
& \mathcal{C}^{\text{v}}_{q_i,q_j}=\frac{1}{2}\left(C^{q_i,q_j} - C^{q_i,\bar q_j}\right)
\end{align}
Here again we have dropped the index ``$k$'' and the dependencies $x$, $z$ and $Q^2$ for readability. In a similar way as in~\eqref{eq:coefDISps}, we can categorise the different contributions according to their electromagnetic charge dependences. Up to $\mathcal{O}(a^2_s)$ the coefficient functions $C^{q_i,q_j}$ can be written as follows:
\begin{align}\label{eq:coefCharge}
&C^{q_i,q_i}=C^{\bar q_i,\bar q_i}=e^2_{q_i} C_{qq}^{\text{NS}}+\frac{1}{N_f}\bigg(\sum_{i} e^2_{q_i}  \bigg) C_{qq}^{\text{PS}}&
&C^{q_i,\bar q_i}=C^{\bar q_i, q_i}=e^2_i (C^1_{q\bar q} - C^2_{q\bar q})&
\nonumber\\
&C^{q_i,q_j}=C^{\bar q_i,\bar q_j}\overset{i\neq j }=e^2_{q_i} C_{qq\prime}^{1}+e^2_{q_j} C_{qq\prime}^{2}+e_{q_i} e_{q_j}  C_{qq\prime}^{3}&
&  C^{q_i,g}= C^{\bar q_i,g}=e^2_{q_i} C_{qg}&
\nonumber\\
&C^{q_i,\bar{q}_j}=C^{\bar q_i,{q}_j}\overset{i\neq j }=e^2_{q_i} C_{qq\prime}^{1}+e^2_{q_j} C_{qq\prime}^{2}-e_{q_i} e_{q_j}  C_{qq\prime}^{3}&
&C^{g,q_i}=C^{g,\bar q_i}=e^2_{q_i} C_{gq} & 
\nonumber\\
& C^{g,g}=\frac{1}{N_f}\bigg(\sum_{i} e^2_{q_i}  \bigg)  C_{gg} \,.& 
\end{align} 
%
At $\mathcal{O}(a_s^0)$ only the $C^{\text{NS}}_{qq}$ differs from zero whereas the gluon-fusion contribution $C_{gq}$ and the gluon-radiation term $C_{qg}$ appear for the first time at NLO. They have been computed for both $F_1$ and $F_L$ up to $\mathcal{O}(a_s)$ in~\cite{Altarelli:1979kv,deFlorian:1997zj}. The remaining coefficients $C_{qq}^{\text{PS}}$, $C_{q\bar q}^{1,2}$, $C_{qq\prime}^{1,2,3}$, $C_{gg}$ in~\eqref{eq:coefCharge} are non-zero for the first time at NNLO and they are generated at this order from the 2 to 3 diagrams of Fig.~\ref{fig:qqNNLO} and~\ref{fig:Diagramsg} \cite{Daleo:2004pn}. Specifically :
\begin{itemize}
\item $C^{(2)}_{gg}$ takes contributions from squaring the diagrams in Fig.~\ref{fig:Diagramsg} and from the squared amplitudes generated by their interferences,
\item $B^2$ in Fig.~\ref{fig:qqNNLO} is the only contribution to $C_{qq}^{\text{PS},\,(2)}$,
\item  $C_{q\bar q}^{1,\,(2)}$ is generated by $A^2$ and $C^2$ with fragmenting anti-quark (quark) of same flavour of the incoming quark (anti-quark) in Fig.~\ref{fig:qqNNLO},
\item $C_{q\bar q}^{2,\,(2)}$ is generated by the interference term $AC$ in Fig.~\ref{fig:qqNNLO} with fragmenting particle being an anti-quark (quark) of same flavour of the incoming quark (anti-quark),
\item $C_{qq\prime}^{1,\,(2)}$ takes contributions only from $C^2$ in Fig.~\ref{fig:qqNNLO} when fragmenting and incoming quark or anti-quark are of different flavours,
\item $A^2$ in Fig.~\ref{fig:qqNNLO} is the only contribution to $C_{qq\prime}^{2,\,(2)}$ when fragmenting and incoming quark or anti-quark are of different flavours,
\item The interference term $AC$ in Fig.~\ref{fig:qqNNLO} contributes to $C_{qq\prime}^{3,\,(2)}$ when fragmenting and incoming quark or anti-quark are of different flavours.
\end{itemize}
Moreover, the $\mathcal{O}(a^2_s)$ contribution to the $C_{qq}^{\text{NS}}$ coefficient generates from loop and radiative corrections to the $\mathcal{O}(a^0_s)$ and $\mathcal{O}(a_s)$  diagrams together with the $A^2$, $C^2$, $D^2$, $AD$, $BC$ contributions form Fig.~\ref{fig:qqNNLO} when the incoming quark (anti-quark) and the fragmenting quark (anti-quark) are of the same flavour. Contributions proportional to  $\sum_i\left(e^2_{q_i}\right)/N_f$ will appear only starting from N${}^3$LO for the coefficients $C^{q_i,g}$ and $C^{g,q_i}$ whereas for $C^{q_i,q_j}$ and $C^{q_i,\bar q_i}$ this will happen at N${}^4$LO. An example of such contributions is given in Fig.~\ref{fig:NmNLO}.

As a last remark of this section, we want to stress the peculiarity of the $C_{qq\prime}^{3}$ coefficient. It generates from diagrams of the type that would vanish in the sum in the inclusive case. In SIDIS however, it isolates the ``valence'' combinations of PDFs and FFs when the incoming and the fragmenting quark or anti-quark are of different flavours. In fact, at NNLO $C_{qq\prime}^{3}$ is the only coefficient that contributes to the last line of Eq.~\eqref{eq:SIDSstrucfunctFlav}.
 %
%

\section{Calculation of the new contributions to the longitudinal structure function}
\label{sec:FL_NNLO}
%

In the last section we have summarized the different NNLO contributions to the structure function that need to be calculated for a full $\mathcal{O}(a_s^2)$ result. In this paper we start by calculating the simplest corrections to the longitudinal structure function, namely the coefficients $C_{L,\,qq\prime}^{1,(2)}$,  $ C_{L,qq\prime}^{2,(2)}$,  $ C_{L,qq\prime}^{3,(2)}$ and $ C_{L,\,gg\prime}^{(2)}$, whose definitions can be found in Eq. \eqref{eq:coefCharge}. As already discussed in Section \ref{sec:NNLO}, they represent two channels that appear for the first time at NNLO: $\gamma^*+q \to q+\bar{q}'+q'$ with fragmenting quark or anti-quark of different flavour of the incoming quark or anti-quark, and $\gamma^*+g \to q+\bar{q}+g$ with the fragmenting parton being the gluon $g$.  From now on, we will indicate these two processes with $qq'$ and $gg$ respectively. Considering only $qq'$ and $gg$ channels, the structure function in \eqref{eq:strucfunc} can be written 
using Eqs.~\eqref{eq:CoefRelation} and \eqref{eq:coefCharge} as
\begin{align}
F^{(2)}_{L,(qq'+gg)}(x,z,Q^2)&=a_s^2(Q^2)\Bigg\{\sum_i^{N_f} e_{q_i}^2\Bigg[
\bigg( q_i^{\text{NS}}+q_S\bigg)\otimes C_{L,qq\prime}^{1,(2)}\otimes \bigg(\sum_{\substack{j\\j\neq i}}^{N_f} \Big( D_{q_j}^{h,\text{NS}} \Big)+(N_f-1)D^h_S\bigg)\nonumber\\[2mm]
&\hspace*{-25mm}+\bigg(\Big(\sum_{\substack{j\\j\neq i}}^{N_f}  q_j^{\text{NS}} \Big)+(N_f-1)q_S\bigg)\otimes C_{L,qq\prime}^{2,(2)}\otimes \bigg( D_{q_i}^{h,\text{NS}}+D^h_S\bigg)+\frac{1}{N_f}g\otimes C_{L,\,gg}^{(2)}\otimes D^h_g\Bigg](x,z) \nonumber\\[2mm]
&\hspace*{-25mm}+\sum_i^{N_f} \sum_{\substack{j\\j\neq i}}^{N_f} e_{q_i}e_{q_j}\Bigg[q^v_i\otimes  C_{L,qq\prime}^{3,(2)} \otimes D^{h,\,v}_{q_j}
\Bigg](x,z)\Bigg\}
\end{align}
Since no lower order diagrams are present for those channels, no loop corrections and no distributions appear at this level of accuracy. This simplifies the calculation considerably. Hereinafter the main highlights of our calculation are presented.

In order to regularize the divergences that appear at the intermediate stages of the computation  we  use dimensional regularization \cite{Bollini:1972ui,'tHooft:1972fi},
i.e., we work in a $d$-dimensional space, with $d=4-2\epsilon$. All quarks are considered massless.

\begin{figure}[t]
	\centering
			\includegraphics[width=12cm]{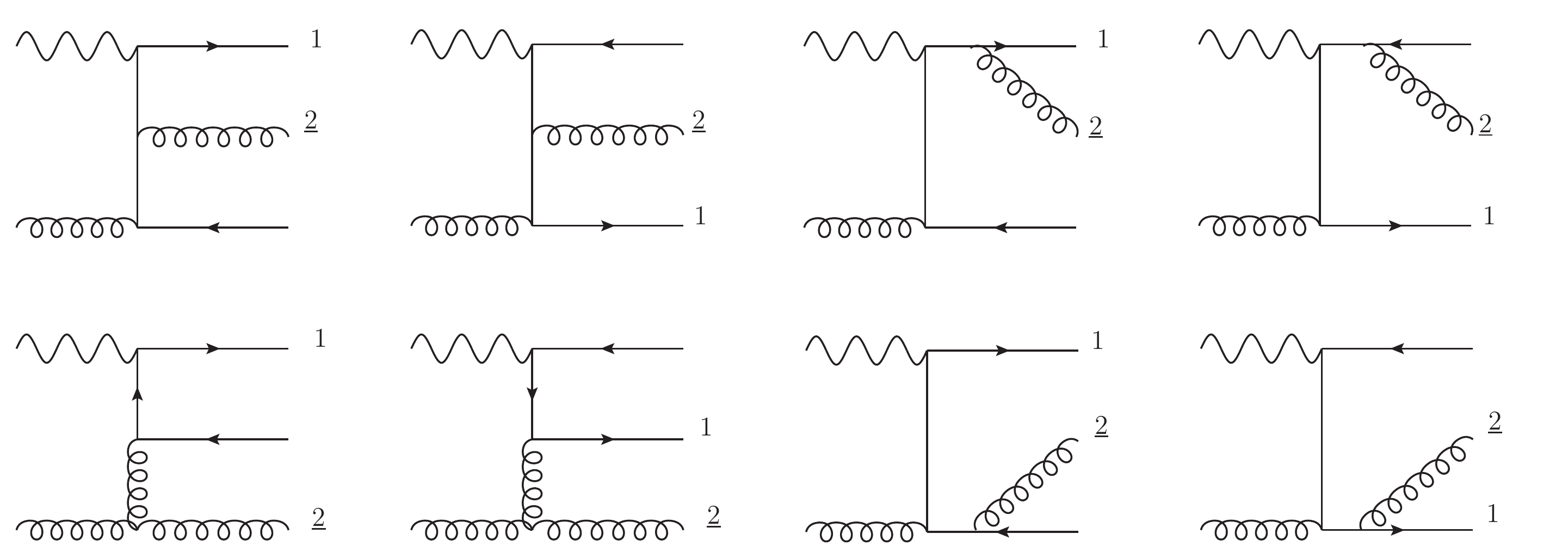}
	\caption{Contributing diagrams to the $gg$ channel ($\gamma^*+g \to q(1)+\bar{q}+``g"(2)$). As for before, particle ``2'' is assumed to be the one fragmenting.}
	\label{fig:Diagramsg}
\end{figure}

The diagrams contributing to $qq'$ and $gg$ channel are shown in Figs.~\ref{fig:qqNNLO} and \ref{fig:Diagramsg}. We compute the squared amplitudes for each channel with the  {\sc Mathematica} packages {\sc FeynArts} \cite{Mertig:1990an} and {\sc FeynCalc} \cite{Hahn:2000kx}. When summing over the gluon helicities we only take into account the physical ones:
\begin{align}
\sum_{\lambda} {\varepsilon}_{\mu}(p,\lambda) {\varepsilon}_{\nu}^*(p,\lambda) = -g_{\mu\nu}+\frac{n_{\mu} \, p_{\nu} + n_{\nu} \, p_{\mu}}{n \cdot p} \, ,
\end{align}
with $n$ an auxiliar vector ($n^2=0$). The explicit dependence on $n$ drops out due to gauge invariance. The same result is obtained by working in a covariant gauge and thus taking external ghosts lines into account.

Since the phase space integration has to be performed over the momenta of the unobserved partons (for instance, quark-antiquark pair for the $gg$ channel), we decide to work in the center of mass frame of these two outgoing partons. In this frame, we still have the chance to choose which one of the remaining momenta  defines the $z$-axis \cite{Beenakker:1988bq}. This choice defines three different sets of kinematic variables. While the set with the photon's momentum ($q$) along the $z$-axis is not useful, since the photon is not massless, the other two sets are convenient for parametrizing different terms of the computation. For all the sets available, we can define
\begin{eqnarray}
2 \, q\, \cdot \, k_h = \frac{Q^2}{x} \left[ 1-x-z-(1-x)(1-z) \, y \right] \, ,
\end{eqnarray}
with $k_h$ being the momentum of the hadronizing parton (gluon in the $gg$ channel and $q'$ or $\bar{q}'$ for the $qq'$ channel), $x$ and $z$ the usual SIDIS variables. 

At the end, the amplitude can be written in terms of just $Q^2$, $x$, $z$, $y$, and the polar and azimuthal angles of the pair of unobserved partons:  $\theta$ and $\phi$ respectively.
Then, we can obtain each one of the coefficients $ C_L^{jk}$ as the finite part of the {\it partonic structure function}, defined by 
\begin{equation}
F_L^{jk}=\frac{1}{4 \pi} \int d\Gamma\,\, {P}_L^{\mu\nu} \,\, \overline{\left| M^{jk} \right|^2}_{\mu\nu} ,
\label{eq:partonicfi}
\end{equation}
where  $d\Gamma$ is the $d$-dimensional phase-space and the longitudinal projector is
\begin{eqnarray}
{P}_L^{\mu\nu}= \frac{8 \, x^2}{Q^2} p^{\mu} \, p^{\nu}  . 
\end{eqnarray}
The $d$-dimensional phase-space can be written as \cite{Matsuura:1988sm}
\begin{eqnarray}
\int d\Gamma & = &\frac{1}{(4 \pi)^{4-2\epsilon}} \, \frac{(s+Q^2)^{1-2\epsilon}}{\Gamma (1-2\epsilon)} \, (1-x)^{1-2\epsilon} \,  z^{-\epsilon} \, (1-z)^{1-2\epsilon}   \nonumber \\ 
	   &   & \times \,\int_0^{\pi} \, d \theta \, \int_0^{\pi} \, d\phi \; (\sin\theta)^{1-2\epsilon} \, (\sin\phi)^{-2\epsilon} \int_0^1 dy  \, \left[ y \, (1-y)\right]^{-\epsilon} \,.
\label{eq:PS}
\end{eqnarray}

All the angular integrals of Eq. \eqref{eq:partonicfi}  can be written, by means of partial fractioning, as
\begin{eqnarray}
I(k,l,a,b,A,B,C)= \int_0^{\pi} \, d \theta \, \int_0^{\pi} \, d\phi \; \frac{ (\sin\theta)^{1-2\epsilon} \, (\sin\phi)^{-2\epsilon}} 
 {(a+b\, \cos\theta)^k (A+B\, \cos\theta + C \, \cos\phi \, \sin\,\theta)^l}.
\label{eq:AngularInt}
\end{eqnarray}
These integrals need to be classified according to the relations their parameters satisfy: i) $a^2=b^2$, ii) $A^2=B^2+C^2$, iii) both relations or iv) neither of them. Besides, the integrals of group ii) can be recasted in terms of those of group i). 
In some cases (in particular whenever an integral of type iv) appears, but also for some integrals of group i)) we can compute the angular integrals in 4 dimensions. Nevertheless, some of the integrals are divergent and we therefore need a $d$-dimensional computation. 
Since the integration over $y$ does not introduce extra poles for the contributions studied in this paper, we can expand the results of the angular integrations up to order 0 in $\epsilon$.

Most of the angular integrals that we need can be found in Appendix C of \cite{Beenakker:1988bq}. We had to compute, however, some unknown ones  that are presented in Appendix \ref{sec:integrals} for the sake of completeness. These new integrals have been computed in 4 dimensions and are valid for groups i) and iv) enumerated above.

Next, we need to perform the integration over $y$, after expanding the integrand up to order 0 in $\epsilon$. 
It is important to notice that this integral is not straightforward. Instead, several changes of variables must be done and some terms must even be rewritten in a clever way to avoid the appearance of spurious divergences in the intermediate steps.

For instance, one of the terms that appear in our computation is 
\begin{eqnarray}
&&\frac{1}{(q-k_h)^2 (q-k_2)^2} = \nonumber \\
&=&\frac{1}{(Q^2+u)} \frac{2}{Q^2} \, I[0, 1,a, b, \frac{Q^2+s-u}{Q^2},\frac{Q^2+s-t-(t+u) \, \cos(\psi)}{Q^2}, -
\frac{(t+u)  \sin (\psi)}{Q^2}] \nonumber \\
&=&\frac{2 \pi  x^2}{Q^4 (1-z)} \frac{\log \left(\frac{(x + z + y (1 - x) (1 - z))+\sqrt{(x + z + y(1 - x) (1 - z))^2 - 4 x z}}{(x + z + y (1 - x) (1 - z))-\sqrt{(x + z + y (1 - x) (1 - z))^2 - 4 x z}}\right)}{(1 + (-1 + x) y)  \sqrt{((x + z + y(1 - x) (1 - z))^2 - 4 x z)}} \,  \nonumber .
\label{eq:example}
\end{eqnarray}
Here, $k_2$ is the momentum of one of the partons in the final state that do not hadronize. 
Written like that, it cannot be integrated. However, after the change of variable ${y \rightarrow \frac{w - x - z}{(1-x)(1-z)}}$,
equation \eqref{eq:example} becomes
\begin{eqnarray}
\frac{1}{(q-k_h)^2 (q-k_2)^2} &=&\frac{2 \pi  x^2}{Q^4} \, \frac{\log (4 x z)-2 \log \left(w+\sqrt{w^2-4 x z} \right)}{(w-x-1) \sqrt{w^2-4 x z}} \, ,
\label{eq:example2}
\end{eqnarray}
whose integral can be performed analytically.

At the end, we obtain the functions $F_L^{jk}$ defined in \eqref{eq:partonicfi}. These contain collinear divergences, that appear as poles in $\epsilon$ (for these processes at NNLO, simple poles in $\epsilon$).  
We factorize these divergences within the  $\overline{\mathrm{MS}}$ scheme, by subtraction of the quantities
\begin{align}
&&\tilde{F}_{L}^{qq'} (x,z) = &\frac{1}{\hat{\epsilon}} \left[  C_{L}^{gq',(1)} (x,z)  \otimes P_{gq}^{(0)} (x) + C_{L}^{qg,(1)} (x,z)  \otimes P_{qg}^{T, (0)} (z)  \right]  , \nonumber \\
&&\tilde{F}_{L}^{gg} (x,z) = &\frac{1}{\hat{\epsilon}} \left[  C_{L}^{qg,(1)} (x,z)  \otimes P_{qg}^{(0)} (x)  + C_{L}^{gq,(1)} (x,z)  \otimes P_{gq}^{T,(0)} (z)  \right]  ,
\label{eq:gi_msbar}
\end{align} 
where $P_{jk}^{(0)}$ are the unpolarized LO splitting functions and  we define
\begin{equation}
\frac{1}{\hat{\epsilon}}=\left[  -\frac{1}{\epsilon} + \gamma_E - \log(4 \pi) \right] , 
\end{equation}
with $\gamma_E=0,5772...$ the Euler constant.
The finite functions obtained after factorization are the coefficients $C_L^{jk}$. Given the length of these coefficients, we do not show them in this paper but they are available upon request.

\section{Results}
\label{sec:Results}
%
We analyze in this section the differences between the semi-inclusive and the inclusive cross section ratio and we show the relevance of the NNLO corrections we have computed. The behavior of the SIDIS ratio $R$ is studied in the range $0.1<z<1$ for different $x$ values. We rely on MSTW PDFs \cite{Martin:2009iq} and DSS fragmentation functions \cite{deFlorian:2007aj}. We fix all scales equal to $Q$ and consider $N_f=4$ active flavours. 

\begin{figure}[ht]
	\centering
			\includegraphics[width=12cm]{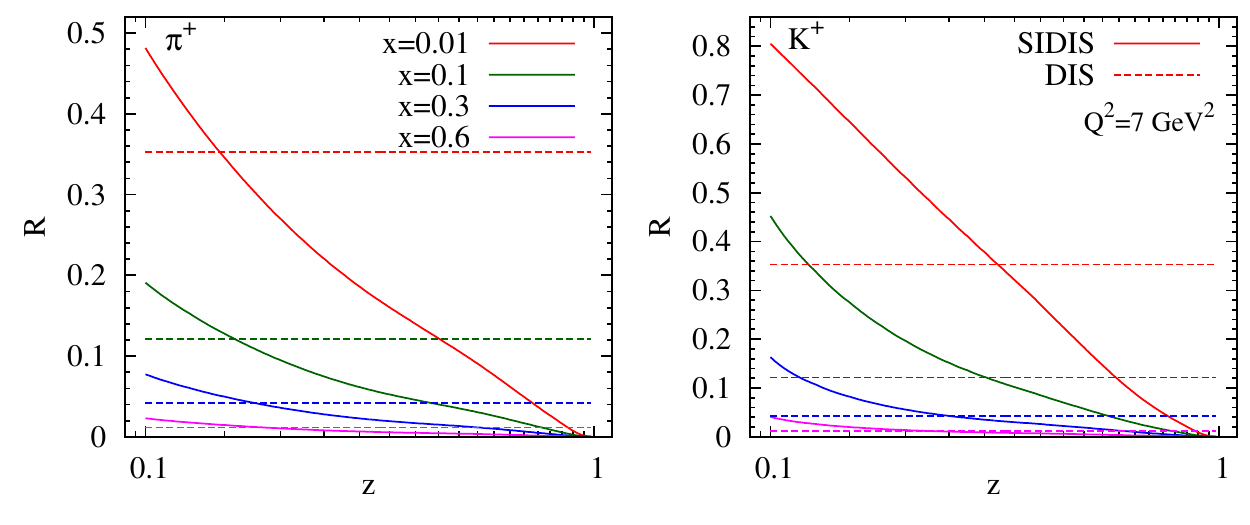}
	\caption{NLO longitudinal-transversal ratio at $Q^2=7 \, {\rm GeV}^2$. The solid curves show the semi-inclusive case, with the observation of a $\pi^+$ (left side) and a $K^+$ (right side) in the final state, while the dashed ones shows the inclusive case.}
	\label{fig:NLO7}
\end{figure}

In Fig \ref{fig:NLO7} we show the semi-inclusive ratio $R$ at NLO for $Q^2=7 \, {\rm GeV}^2$ when a $\pi^+$ (left) or a $K^+$ (right) are observed in the final state. We compare it with the value of the fully-inclusive ratio also at NLO (dashed line), which does not depend on $z$. As we can see, the fully-inclusive and the semi-inclusive results may differ by a factor of two in the relevant kinematical region (and even more close to the edges). Thus, an accurate semi-inclusive description of $R$ is crucial for phenomenological analyses and may not be, in general, approximated by the inclusive one.

\begin{figure}[ht]
	\centering
			\includegraphics[width=14cm]{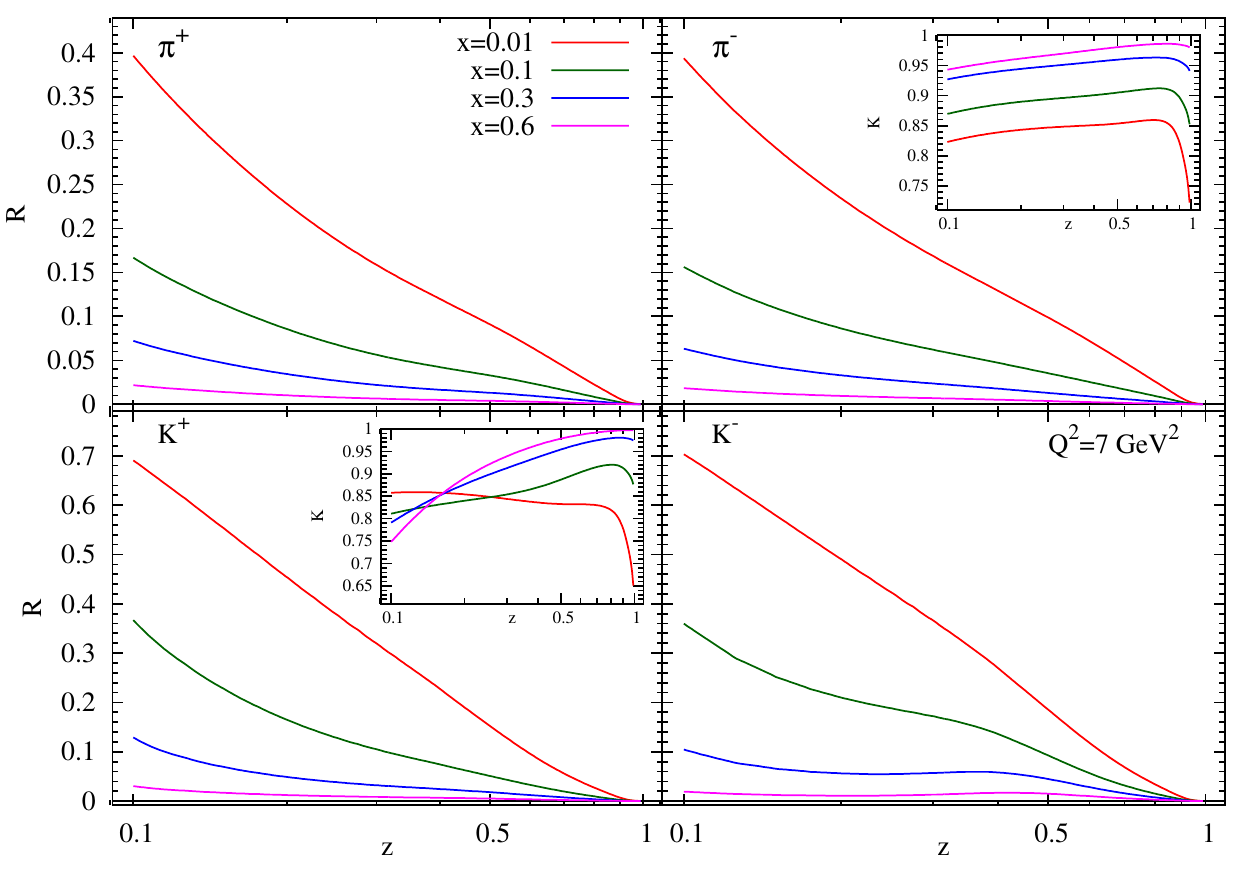}
	\caption{Longitudinal-transversal ratio computed taking into account the NNLO contributions considered in this paper for different hadrons observed in the final state, at  $Q^2=7 \, {\rm GeV}^2$.  The inset plots show the ratio between the NNLO and the NLO computation presented in this paper.}
	\label{fig:NNLO7}
\end{figure}

In Fig. \ref{fig:NNLO7} we present the predictions for the semi-inclusive $R$ ratio including the contributions to the longitudinal structure function at NNLO considered in our work, at $Q^2= 7 \, {\rm GeV}^2$ and for different final-state hadrons. We should mention that NLO PDFs and FFs are used in order to fully  appreciate the effect of the corrections introduced by the new coefficients.The inset plots show the ratio between the NNLO and the NLO computation presented in this paper.

We can see that the correction introduced by the NNLO contributions studied in this paper turn out to be negative and, therefore, tend to considerably reduce the value of $R$ with respect to the previous order. The corrections are specially sizeable for the low-$z$ and high-$z$ regions. 
This is likely due to the appearance of logarithmic terms introduced by the NNLO contributions and therefore only present in the numerator of the ratio $R$.

Due to the quark composition of kaons and protons, the contribution coming from $F^{(2)}_{L,qq'}$ is dominant for $K^+$ and $K^-$, for all $x$ values analyzed. In fact, more than 80$\%$ of the new NNLO correction arises from that channel. The situation is different for pions and in this case the corrections from $qq'$ and $gg$ channels turn out to be even of the same order for some kinematic regions.

Despite of including only a subset of contributions at NNLO, those in principle expected to be small due to their particular structure, the corrections to the longitudinal structure function turn out to be rather sizable, making the calculation of the full corrections even more mandatory.

\section{Summary and Outlook}
\label{sec:Conclusions}
%
We have presented a first calculation of the $\mathcal{O}(\alpha_s^2)$ contributions to the SIDIS longitudinal structure function generated by partonic channels appearing for the first time at NNLO, together with an extensive discussion of the general aspects useful for the organization of the complete NNLO calculation. 

We have started by studying the flavour decomposition of the full cross section. We have shown how to express both longitudinal and transversal structure functions in terms of usual singlet and non-singlet combinations of PDFs and FFs relevant in global analysis fits, thus exposing the flavour structure of the SIDIS cross section. For instance, one can notice that a specific contribution calculated here in this paper, namely $C_{q q'}^3(2)$, isolates a particular ``valence'' flavor contribution, to which a NLO cross section would be insensitive.
Along the way, we have given a summary of the different diagrammatic contributions to the partonic sub-processes in view of the full NNLO calculation. In the same spirit, a general recursive formula to reconstruct the dependence on both factorization scales involved in the SIDIS  process at an arbitrary order in the strong coupling constant was derived. 

In a second part, we have given details of our computation for the two channels of interest in this paper. In particular, the procedure to analytically calculate the phase space integrals has been discussed and a set of new angular integrals was presented. 
Although the calculated channels are a partial component of the full set of NNLO corrections to the cross section, we have discussed some phenomenological study done on the observable $R$ as a theoretical investigation of the relevance of NNLO corrections to the cross section. It turns out that the small fraction of contributions calculated so far do exhibit sizeable corrections especially in the low-$z$ and high-$z$ region, likely due to the appearance of logarithms that are only present in the numerator. 

Not only these observations are of theoretical interest, but stating whether or not this is also the case once all corrections up to NNLO are added is of great importance when it comes to extracting FFs from SIDIS data. As it is well know, including SIDIS data in a global analysis helps disentangling the different single flavour contributions to the fragmenting process. Even more at NNLO where the appearance of new channels discriminates specific new combinations of PDFs and FFs. Nonetheless, only with the full calculation available one can at the end asses their phenomenological relevance 
in the overall picture of a global analysis. 
With the precision of the FFs having been recently extended to NNLO and beyond in the context of electron-positron to pion only analysis, it is then natural to try to acquire the sort of ``know how'' needed to complete a NNLO calculation of the SIDIS process. 
In an attempt to attack this problem with an analytical approach, we have started by computing the first simplest corrections to the longitudinal structure function as a playground where to explore and organize the future complete calculation. From a theoretical point of view, only with analytical results available one gathers useful insight in the structure of the perturbative series. Our intention is to proceed on this path and complete the full calculation analytically. This could be relevant for further applications which go beyond PDFs and FFs analyses. For example, knowing the structure of sub-leading logarithms connected with precise phase-space configurations is an essential ingredient in order to extend resummation techniques to higher accuracy.

\section*{Acknowledgments}
%
We are very grateful to Werner Vogelsang for many valuable contributions and discussions throughout this work.
D.P.A.\ acknowledges partial support from the Fondazione Cassa Rurale di Trento.
D.P.A.\ was supported by the Deutsche Forschungsgemeinschaft (DFG) under grant no. VO 1049/1.

\appendix

\section{Reconstruction of scales\label{sec:scales}}

In order to reconstruct the full dependences on the factorization scales $\mu_F$ and $\mu_I$ at every order in perturbation theory, one can use a renormalization group approach similar to what was done in~\cite{vanNeerven:2000uj} for the totally inclusive DIS case or in~\cite{Anderle:2016czy} for the semi-inclusive electron-positron annihilation (SIA). In~\cite{Anderle:2016czy}, an alternative method based on the mass factorization procedure was discussed in order to obtain the same results. We have extended both methods to the SIDIS case and found full agreement between them. 

Hereinafter, we will review the extension of the renormalization group approach method to the SIDIS case in order to present a general recursive formula which can be utilize to reconstruct the scale dependence on the two factorization scales $\mu_I$ and $\mu_F$ at an arbitrary order in the strong coupling constant $a_s=\alpha_s/4\pi$. We are going to show the calculation only for the first term in Eq.~\ref{eq:SIDSstrucfunctFlav} since it is the most complicated case due to its matrix structure. To simplify the calculation we set the renormalization scale $\mu_r=\mu_F$ but keep $\mu_F\neq\mu_I\neq Q^2$. The reintroduction of the renormalization scale dependence can be easily achieved by re-expanding the result expressed as a function of $a_s(\mu_F^2)$ in terms of $a_s(\mu_r^2)$. The third order expansion of $a_s$ reads~\cite{Moch:2005id}
\begin{align}\label{eq:expansion}
a_s(\mu^2)&=\frac{a_s(\mu_0^2)}{X(\mu^2)} - \frac{a_s^2(\mu_0^2)}{X^2(\mu^2)} \left(\frac{\beta_1}{\beta_0}\log X(\mu^2)\right)\nonumber\\
&+\frac{a_s^3(\mu_0^2)}{X^3(\mu^2)}
\left( \frac{\beta_1^2}{\beta_0^2} \Big( \log^2X(\mu^2)-\log X(\mu^2)- 1 + X(\mu^2)    \Big) + \frac{\beta_2}{\beta_0}\Big(1-X(\mu^2)\Big)\right)+\dots\,.
\end{align}
where $X(\mu^2)=1-\beta_0\log({\mu_0^2/\mu^2})$, and $\mu_0$ is a reference scale that in our case corresponds to $\mu_r$.\\
We denote
\begin{align}
F^S_k(\mu_I^2,\mu_F^2)
&=(q_S,  g )(\mu^2_I)\otimes \Bigg(\begin{matrix}  \mathcal{C}^{S,D_S}_k& \mathcal{C}^{S,g}_k \\[2mm] \mathcal{C}^{g,D_S}_k& \mathcal{C}^{g,g}_k\end{matrix}\Bigg)\left(a_s(\mu^2_F),L_I,L_M\right)\otimes \Bigg(\begin{matrix}  D^h_S \\[2mm] D^h_g\end{matrix}\Bigg)(\mu^2_F)\nonumber\\[2mm]
&=\bs{q}(\mu^2_I)\otimes \bs{  \mathcal{C}}^{S}_{k}\left(a_s(\mu^2_F),L_I,L_M\right)\otimes\bs{D}^h_S(\mu_F^2)\;,
\end{align}
where $k\in \{1,\,L\}$, $L_I=\log({Q^2/\mu_I^2})$, $L_F=\log({Q^2/\mu_F^2})$ and the dependences on $x$ and $z$ were dropped for clarity in the notation. By taking the double Mellin transformation of the previous equation, we can further simplify the calculation. Since convolutions between functions are represented in Mellin space by simple multiplications between moments of them, we can then write
\begin{align}\label{Melmo}
\tilde{F}^S_k(N,M,\mu_I^2,\mu_F^2,Q^2)=&\bs{\tilde{ q}}^N(\mu_I^2)\times\bs{\tilde{{\cal C}}}^S_{k}\left(N,M,\alpha_s(\mu_F^2),L_I,L_F\right)\times\bs{\tilde{D}}_{S}^{h,M}(\mu_F^2),
\end{align}
where the symbol $\times$ denotes the standard matrix multiplication and 
\begin{align}
\tilde{F}^S_k(N,M,\mu_I^2,\mu_F^2,Q^2)&\equiv\int_0^1 dx x^{N-1}\int_0^1 dz z^{M-1}F^S_k(x,z,\mu_I^2,\mu_F^2,Q^2)\nonumber\\[2mm]
\bs{\tilde{q}}^N(\mu_I^2)&\equiv\int_0^1 dx x^{N-1}\bs{q}(x,\mu^2_I),\nonumber\\[2mm]
\bs{\tilde{D}}_{S}^{h,M}(\mu_F^2)&\equiv\int_0^1 dz z^{M-1}\bs{D}^h_{S}(z,\mu_F^2),\nonumber\\[2mm]
\bs{\tilde{{\cal C}}}^S_{k}\left(N,M,a_s(\mu_F^2),L_I,L_F\right)&\equiv\int_0^1 d\hat{x}\hat{x}^{N-1}\int_0^1 d\hat{z}\hat{z}^{M-1}\bs{\cal{C}}^S_{k}\left(\hat{x},\hat{z},a_s(\mu_F^2),L_I,L_F\right) .\nonumber\\[2mm]
\end{align}

The dependence of each entry of the matrix  $\bs{\mathcal{\tilde{C}}}^{S}_{k}$ on the factorization scales $\mu_I$ and $\mu_F$ can be expressed as 
\begin{align}
\label{eq:coefscales}
\mathcal{\tilde{C}}_{k,ij}^{\text{S}}\left(a_s(\mu_F^2),L_I,L_F\right) =& \sum_{n=0}^\infty a_s^n(\mu_F^2)  \left(  \tilde{c}_{k,ij}^{(n,0,0)} + \sum_{\kappa=1}^{n}
\tilde{c}_{k,ij}^{(n,\kappa,0)} L_I^\kappa + \sum_{l=1}^{n}
\tilde{c}_{k,ij}^{(n,0,l)} L_F^l  \right.\nonumber \\ 
&\left.+\sum_{\kappa=1}^{n} \sum_{l=1}^{n-\kappa} \tilde{c}_{k,ij}^{(n,\kappa,l)} L_I^\kappa L_F^l  \right)\,,
\end{align}
The coefficients $\tilde{c}_{k,ij}^{(n,0,0)}$ are the direct result of the perturbative calculation with $\mu_F^2=\mu_I^2=\mu_r^2=Q^2$ while the $\tilde{c}_{k,ij}^{(n,\kappa,0)}$, $\tilde{c}_{k,ij}^{(n,0,l)}$, $\tilde{c}_{k,ij}^{(n,\kappa,l)}$ can be calculated order by order
in $a_s$ solving the renormalization group equations (RGEs) for the fatorization scales. They follow directly from the request that $\frac{\partial}{\partial\log\mu_I^2} \tilde{F}^S_k \overset{!}{=} 0$ and $\frac{\partial}{\partial\log\mu_F^2} \tilde{F}^S_k \overset{!}{=} 0$ and they read
\begin{align}\label{eq:RGE1}
&\bigg(\Big[\frac{\partial}{\partial \log\mu_I^2}\Big]\delta_{im}  + \tilde{P}^{Transp}_{im}(N,\mu_I^2) \bigg)\mathcal{C}^S_{k,mj}(N,M,a_s(\mu_F),L_I,L_F) = 0
\\[5mm]
\label{eq:RGE2}
&\bigg(\Big[\frac{\partial}{\partial \log\mu_F^2} +  \beta(a_s)\frac{\partial}{\partial a_s}\Big]\delta_{mj}  + \tilde{P}^{T}_{mj}(M,\mu_F^2) \bigg)\mathcal{C}^S_{k,im}(N,M,a_s(\mu_F),L_I,L_F) = 0\,.
\end{align}
Here $\tilde{P}^{Transp}_{im}(N,\mu_I^2)$ corresponds to the $im$ entry of the matrix resulting from the trasposition of
\begin{equation}\label{eq:PTmatrix}
\bs{\tilde{P}} (N,\mu_I^2) \equiv \sum_{i=0}^\infty a_s^{i+1} \bs {\tilde{P}}^{(i)}(N,\mu_I^2) \equiv \sum_{i=0}^\infty a_s^{i+1}  \Bigg (\begin{matrix}\tilde{P}^{(i)}_{qq} & \tilde{P}^{(i)}_{gq}\\[2mm] \tilde{P}^{(i)}_{qg} & \tilde{P}^{(i)}_{gg} \end{matrix} \Bigg )(N,\mu_I^2)
\end{equation}
defined as the single Mellin transform of the matrix appearing in the first equation of~\eqref{eq:DGLAPsinglet}. On the other side, $\tilde{P}^T_{mj}(M,\mu_I^2)$ represents the $mj$ entry of the time-like $\bs{\tilde{P}}^T (M,\mu_I^2)$ matrix defined as the single Mellin transform of the matrix appearing in the second equation of~\eqref{eq:DGLAPsinglet}.\\
Inserting Eq.~\eqref{eq:coefscales} in~\eqref{eq:RGE1} and~\eqref{eq:RGE2} one is left with a system of linear equations in the coefficients $\tilde{c}_{k,ij}^{(n,\kappa,0)}$, $\tilde{c}_{k,ij}^{(n,0,l)}$ and $\tilde{c}_{k,ij}^{(n,\kappa,l)}$ which can be solved recursively order by order in $a_s$ for every fixed value of $\kappa$ and $l$ once the results for $\kappa=0$ and $l=0$ are given. If we define $\bs{\tilde{c}}_{k}^{(n,\kappa,l)}$  to be the matrix with entries $\tilde{c}_{k,ij}^{(n,\kappa,l)}$, the formal solution for a fixed order $\mathcal{O}(a_s^n)$ can be recursively written as
\begin{align}
\label{eq:recursivel0}
\bs{\tilde{c}}_{k}^{(n,\kappa,0)}&\overset{\kappa\neq0}= \frac{1}{\kappa}
\sum_{w=\kappa-1}^{n-1}\bs{\tilde{P}}^{(n-w-1),\,Transp}\times \bs{\tilde{c}}_{k}^{(w,\kappa-1,0)}\nonumber\\[2mm]
&+\frac{1}{\kappa}\sum_{p=0}^{n-2}\sum_{q=0}^{\kappa-2}\;\left(\sum_{i=0}^{n-p-\kappa+q} A^{n-p}_{i,\;\kappa-q-1}\bs{\tilde{P}}^{(i),\,Transp}\right)\times\bs{\tilde{c}}_{k}^{(p,q,0)}\\[5mm]
\label{eq:recursivelk}
\bs{\tilde{c}}_{k}^{(n,\kappa,l)}&\overset{l\neq0}= \frac{1}{l}\sum_{j=l-1+\kappa}^{n-1} \bs{\tilde{c}}_{k}^{(j,\kappa,l-1)}\times
\left(\bs{\tilde{P}}^{T,(n-1-j)}-{\mbox{\boldmath $1$}}(j \,\beta_{n-1-j})\right)\;,
\end{align}
where all dependences have been dropped to simplify the notation. The coefficients $\bs{\tilde{c}}_{k}^{(n,0,l)}$ are also given by the formula~\eqref{eq:recursivelk}. All terms $\bs{\tilde{c}}_{k}^{(n,\kappa,l)}$ with $\kappa+l>n$ recursively generated by the above equations are obviously set to be equal zero. The coefficient $A^{n-p}_{i,\;\kappa-q-1}$, introduced in Eq.~\eqref{eq:Pexpansion}, appears in the last line of Eq.~\eqref{eq:recursivel0} since the space-like splitting functions showing in~\eqref{eq:RGE1} are given as a function of $\mu_I$. Nonetheless, one has to take great care when solving the system of equations~\eqref{eq:RGE1} and~\eqref{eq:RGE2} and re-expand $\tilde{P}^{Transp}_{im}(N,\mu_I)$ around the same $a_s(\mu_F)$ consistently with the one chosen in Eq.~\eqref{eq:coefscales}. As a consequence, Eqs.~\eqref{eq:recursivel0} and~\eqref{eq:recursivelk} can be correct only up to $\beta_i$ terms neglected in the expansion of $a_s$ (see Eq.~\eqref{eq:expansion}).
To regain the expressions in the $(x,z)$ space one has to formally perform a double Mellin inverse 
\begin{equation}\label{eq:inverse}
\bs{\cal{C}}^S_{k}\left(\hat{x},\hat{z},a_s(\mu_F^2),L_I,L_F\right)= \int_{{\cal C}_N}
\frac{d N}{2\pi i}\hat{x}^{-N}  \int_{{\cal C}_M}
\frac{d M}{2\pi i}\hat{z}^{-M}
\bs{\tilde{{\cal C}}}^S_{k}\left(N,M,a_s(\mu_F^2),L_I,L_F\right)\;,
\end{equation}
where ${\cal C}_N$ and ${\cal C}_M$ are contour chosen in the $N$ and $M$ complex moment space respectively.
Assuming that for a fixed order $\mathcal{O}(a_s^n)$ the coefficient $\bs{\tilde{{\cal C}}}^{S,(n)}_{k}$ is integrable along the contours ${\cal C}_N$ and ${\cal C}_M$, we have that
\begin{equation}\label{eq:inversen}
\bs{c}_{k}^{(n,\kappa,l)}(\hat{x},\hat{z})= \int_{{\cal C}_N}
\frac{d N}{2\pi i}\hat{x}^{-N}  \int_{{\cal C}_M}
\frac{d M}{2\pi i}\hat{z}^{-M}
\bs{\tilde{c}}_{k}^{(n,\kappa,l)}(N,M)\;
\end{equation}
and the expressions~\eqref{eq:recursivel0} and~\eqref{eq:recursivelk} can be translated for the $\bs{c}_{k}^{(n,\kappa,l)}$ coefficients by symbolically dropping the ``$\sim$'' and substituting ``$\times$'' with ``$\otimes$''.\\

This procedure can be easily extended for the remaining lines of Eq.~\eqref{eq:SIDSstrucfunctFlav} by simply substituting the matrices $\bs{\tilde{P}}^{T,(i)}$ and $\bs{\tilde{P}}^{(i)}$ with the corresponding ``non-singlet'' scalar function $\tilde{P}^{T,+\,,(i)}$, $\tilde{P}^{T,v\,,(i)}$, $\tilde{P}^{+\,,(i)}$ or $\tilde{P}^{v\,,(i)}$ in~\eqref{eq:recursivel0} and~\eqref{eq:recursivelk}. Here the symbol ``$\sim$'' denotes as before the single Mellin moment of the corresponding function. At the same time $\bs{\tilde{c}}_{k}^{(n,\kappa,l)}$ will represent each time a scalar function, a vector or a transposed vector accordingly to how the coefficient functions appear in~\eqref{eq:SIDSstrucfunctFlav}.

\section{Integrals \label{sec:integrals}}

We show below the results of some angular integrals needed for the computation of the SIDIS longitudinal structure functions, which are not available in the literature. The subscript $4$ indicates that they have been computed in 4-dimensions. Except for the first three integrals (valid for every set of parameters) the results are only valid for $A^2 \neq B^2+C^2$.

\begin{flalign}
I_4[-4,0]&=2 \pi  \left(a^4+2 a^2 b^2+\frac{b^4}{5}\right) \\
I_4[-3,0]&=2 \pi  a \left(a^2+b^2\right)    \\
I_4[-2,-1]&=\frac{2}{3} \left(3 \pi  a^2 A+2 \pi  a b B+\pi  A b^2\right) \\
I_4[-4,2]&=
  \frac{\pi}{3 \left(B^2+C^2\right)^4 \left(A^2-B^2-C^2\right)}  \nonumber \\
    &\times \left\{A^2 b^2 \left(B^2+C^2\right)   \left[36 a^2 \left(2  B^4+B^2 C^2-C^4\right)+b^2 \left(-16 B^4+84 B^2 C^2-15 C^4\right)\right]  \right.\nonumber \\
    &- 12 a A b B \left(B^2+C^2\right)^2 \left[2 a^2 \left(B^2+C^2\right)+b^2 \left(9 C^2-4 B^2\right)\right] \nonumber \\
    &+ 2 \left(B^2+C^2\right)^2 \left[3 a^4 \left(B^2+C^2\right)^2-18 a^2 b^2 \left(B^4-C^4\right)-b^4 \left(B^4+6 B^2 C^2-3 C^4\right)\right] \nonumber \\
    &- \left. 36 a A^3 b^3 B \left(2 B^4-B^2 C^2-3 C^4\right)+3 A^4 b^4 \left(8 B^4-24 B^2 C^2+3 C^4\right)\right\} \nonumber \\
  &+\frac{\pi  b}{2 \left(B^2+C^2\right)^{9/2}} \log \left(\frac{A+\sqrt{B^2+C^2}}{A-\sqrt{B^2+C^2}}\right) \nonumber \\
   &\times \left\{-3 A b \left(B^2+C^2\right) \left(4 a^2 \left(2   B^4+B^2 C^2-C^4\right)-b^2 C^2 \left(C^2-4 B^2\right)\right) \right. \nonumber \\
   &+ 4 a B \left(B^2+C^2\right)^2 \left(2 a^2  \left(B^2+C^2\right)+3 b^2 C^2\right) + 12 a A^2 b^2 B \left(2 B^4-B^2 C^2-3 C^4\right) \nonumber \\
   &+ \left. A^3 b^3 \left(-8 B^4+24 B^2  C^2-3 C^4\right)\right\}  \\
I_4[-4,1]&= 
   -\frac{\pi  b} {12 \left(B^2+C^2\right)^4} \nonumber \\
   &\times \left\{A b \left(B^2+C^2\right) \left[72 a^2 \left(2  B^4+B^2 C^2-C^4\right)+b^2 \left(8 B^4+48 B^2 C^2-15 C^4\right)\right]  \right. \nonumber \\
   &-32 a B \left(B^2+C^2\right)^2 \left[3 a^2 \left(B^2+C^2\right)+b^2 \left(B^2+3 C^2\right)\right] \nonumber \\
   &-48 a A^2 b^2 B \left(2 B^4-B^2 C^2-3 C^4\right) \nonumber \\
   &+ \left. 3 A^3 b^3 \left(8 B^4-24 B^2 C^2+3 C^4\right)\right\} \nonumber \\
   &+\frac{\pi} {8 \left(B^2+C^2\right)^{9/2}} \log \left(\frac{A+\sqrt{B^2+C^2}}{A-\sqrt{B^2+C^2}}\right) \nonumber \\
   &\times \left\{6 A^2 b^2 \left(B^2+C^2\right) \left(4 a^2 \left(2 B^4+B^2 C^2-C^4\right)-b^2 C^2 \left(C^2-4 B^2\right)\right)  \right.\nonumber \\
   &-16 a A b B \left(B^2+C^2\right)^2 \left(2 a^2 \left(B^2+C^2\right)+3 b^2 C^2\right)  \nonumber \\
   &+\left(B^2+C^2\right)^2 \left(8 a^4 \left(B^2+C^2\right)^2  + 24 a^2 b^2 C^2 \left(B^2+C^2\right)+3 b^4 C^4\right) \nonumber \\
   &+16 a A^3 b^3 B \left(-2 B^4+B^2 C^2+3 C^4\right)  \nonumber \\
   &+ \left. A^4 b^4 \left(8 B^4-24 B^2 C^2+3 C^4\right)\right\}  
\end{flalign}
\begin{flalign}
I_4[-3,2]&= \frac{\pi}{\left(B^2+C^2\right)^3 \left(A^2-B^2-C^2\right)} \nonumber \\
   & \times \left\{2 a^3 B^6+6 a^3 B^4 C^2+6 a^3 B^2 C^4+2 a^3 C^6-6 a^2  A b B^5 \right. \nonumber \\
   &-12 a^2 A b B^3 C^2-6 a^2 A b B C^4 +12 a A^2 b^2 B^4+6 a A^2 b^2 B^2 C^2-6 a A^2 b^2 C^4-6  a b^2 B^6  \nonumber \\
   &-   6 a b^2 B^4 C^2 + 6 a b^2 B^2 C^4 + 6 a b^2  C^6-6 A^3 b^3 B^3+9 A^3 b^3 B C^2+4 A b^3 B^5\nonumber \\
   &- \left. 5 A  b^3 B^3 C^2-9 A b^3 B  C^4\right\}  \nonumber  \\
   &+\frac{3 \pi  b} {2 \left(B^2+C^2\right)^{7/2}} \log  \left(\frac{A+\sqrt{B^2+C^2}}{A-\sqrt{B^2+C^2}}\right) \nonumber \\
   & \times \left\{2 a^2 B^5+4 a^2 B^3 C^2+2 a^2 B C^4 -4 a A b B^4-2 a A b B^2 C^2+2 a A b C^4 \right. \nonumber \\
   &+ \left. 2 A^2 b^2  B^3-3 A^2 b^2 B C^2+b^2 B^3 C^2+b^2 B C^4\right\} \\
I_4[-3,1]&=\frac{\pi  b}{3\left(B^2+C^2\right)^3} \nonumber \\
   & \times\left\{18 a^2 B^5+36 a^2 B^3 C^2+18 a^2 B C^4-18 a A b B^4-9 a A b B^2 C^2+9 a A b C^4 \right. \nonumber \\
   &+ \left. 6 A^2 b^2 B^3-9 A^2 b^2 B C^2+2 b^2 B^5+8 b^2   B^3 C^2+6 b^2 B C^4\right\} \nonumber \\
   &-\frac{\pi  \left(-a B^2-a C^2+A b B\right)}{2   \left(B^2+C^2\right)^{7/2}}  \log \left(\frac{A+\sqrt{B^2+C^2}}{A-\sqrt{B^2+C^2}}\right) \nonumber \\
   &\times \left\{2 a^2 B^4+4   a^2 B^2 C^2+2 a^2 C^4-4 a A b B^3 \right. \nonumber \\
   &- \left. 4 a A b B   C^2+2 A^2 b^2 B^2-3 A^2 b^2 C^2+3 b^2 B^2   C^2+3 b^2 C^4\right\}
\end{flalign}


%
\end{document}